\documentclass[preprint,11pt,authoryear, nonatbib]{elsarticle}
\usepackage{graphicx}
\usepackage[natbibapa]{apacite}

\usepackage{booktabs}
\usepackage[export]{adjustbox}
\usepackage{graphicx}
\usepackage{graphbox}
\usepackage{colortbl}
\usepackage{subcaption}
\usepackage{enumitem}
\usepackage[natbibapa]{apacite}

\usepackage{amsmath}
\usepackage{amssymb}
\usepackage{lineno}
\usepackage[hidelinks]{hyperref}
\usepackage[utf8]{inputenc}
\usepackage{newunicodechar}
\usepackage[utf8]{inputenc}
\usepackage{textcomp}

\begin{document}

\begin{frontmatter}
\title{The Body as Status: Muscularity, Engagement, and Body Image Risk on \#GymTok}

\author[1]{Magdalayna Curry}
\ead{magcurry@usc.edu}
\author[2]{Minh Duc Chu\corref{cor1}}
\ead{mhchu@usc.edu}
\author[1]{Changhao Yan}
\ead{cyan5417@usc.edu}
\author[3]{Stuart B. Murray\fnref{ci}}
\ead{sbmurray@mednet.ucla.edu}
\author[2]{Kristina Lerman\fnref{ci}}
\ead{lerman@isi.edu}
\author[1]{Lindsay E. Young}
\ead{lindsay.young@usc.edu}

\address[1]{Annenberg School for Communication and Journalism, University of Southern California, Los Angeles, CA, USA}
\address[2]{Information Sciences Institute, University of Southern California, Marina del Rey, CA 90292, USA}
\address[3]{Department of Psychiatry and Biobehavioral Sciences, University of California, Los Angeles, CA 90095, USA}

\fntext[ci]{Stuart B. Murray and Kristina Lerman have offered expert witness testimony on matters relating to the topic of this paper.}

\cortext[cor1]{Corresponding author: Minh Duc Chu, Information Sciences Institute, University of Southern California, Marina del Rey, CA 90292, USA. Email: mhchu@usc.edu}

\begin{abstract}
Body image concerns among boys and young men are increasingly oriented toward muscularity, with social media serving as a central context for communicating and evaluating these ideals. While prior research has focused on the thin-ideal, less is known about how the muscular-ideal is represented and reinforced on visual social media platforms. This study examines (1) dominant content themes, (2) perceived harm to body image, and (3) engagement patterns across \#GymTok, a muscularity-oriented fitness subculture on TikTok.
We conducted a content analysis of 2,210 \#GymTok videos annotated by clinical experts across themes like self-objectification, rigid dieting, excessive exercise, supplement and steroid use, and masculinity. Annotators also rated the perceived harm of videos to the viewers' body image, and depicted bodies were coded according to muscularity level. Perceived harm varied across content themes, with supplement- and steroid-related content rated as most harmful. Engagement was positively associated with both muscularity and perceived harm: videos depicting more muscular bodies and those rated as more harmful received greater views, likes, shares, and comments. Although less prevalent, masculinity-focused content generated the highest engagement. These findings suggest that TikTok may not only expose users to muscular ideals and potentially harmful behaviors, but also algorithmically amplify them. By increasing the visibility of highly muscular and harmful content, recommendation systems may intensify social comparison processes, while objectification elevates the muscular body into a marker of status, masculinity, and social worth. Together, these dynamics may contribute to body image risk among boys and young men.

\end{abstract}


\end{frontmatter}

\section{Introduction}
Evolving conceptions of masculinity are increasingly shaped by online culture, with growing scholarly and public attention directed toward how digital environments influence sense of identity, self-worth, and body image concerns in boys and men \citep{bell2025active,potts2023behind}. Contemporary discourse surrounding the “male loneliness epidemic,” the rise of the “manosphere,” and appearance-focused trends such as “looksmaxxing” reflects broader concerns about how young men construct masculinity in the age of highly visible, algorithmically mediated social media platforms \citep{ribeiro2021evolutionmanosphereweb, habib2022making}. Within these spaces, masculine identity is increasingly negotiated through the body, as appearance-based hierarchies and peer evaluation of physical traits promote the internalization of idealized body standards and encourage extreme self-optimization practices as a means of attaining and signaling masculine status  \citep{halpin2025help, sousbois2025incels}. These pressures are particularly salient for Gen Z boys and young men, for whom social media serves as a primary context for identity formation, social comparison, and the internalization of masculine norms~\citep{dane2023social}. 

A pervasive source of adolescent boys' and young mens' notions of masculinity is fitness-related social media content. A nationally representative survey of adolescent boys age 11 to 17 conducted by Common Sense Media \citep{robb2025boys} revealed that 91 percent of boys encounter body transformation messages online, with about one in four reporting that this content makes them feel pressure to change the way they look. Studies have consistently linked exposure to fitness-related social media content, such as images of physically fit peers or influencers performing exercises, promoting supplements, or advocating bodybuilding lifestyles, with body dissatisfaction and negative mental health outcomes, including increased suicidal ideation and eating disorder symptomatology \citep{holland2017strong, tennfjord2025social, lepesheva2021fitness}. Yet much of this research has focused on women's experiences with thin-ideal content, establishing that exposure to fitness posts and fitspiration imagery (i.e., fitness-oriented social media content intended to motivate exercise and healthy behaviors \citep{tiggemann2018strong}) decreases body esteem, increases body dissatisfaction, and reinforces internalization of idealized appearance standards \citep{fioravanti2021fitspiration, yan2025your, barnes2023comparison}. Comparatively, less attention has been given to how young men engage with muscularity-oriented content, despite evidence that men are more likely to actively seek, produce, and interact with such content as a means of self-improvement and status signaling~\citep{mayoh2021young}. In such environments, repeated exposure to highly muscular bodies may normalize extreme physiques and intensify upward social comparisons, increasing body dissatisfaction and risks of muscle dysmorphia, disordered eating, and appearance- and performance-enhancing drug (APED) use~\citep{hilkens2021social,ganson2025associations}. 

TikTok represents an especially consequential platform for these processes. TikTok's fitness-oriented subculture---colloquially referred to as “Gym\-Tok”---represents a crucial yet understudied context for examining how muscularity, discipline, and appearance norms are communicated and reinforced among a predominantly young adult audience\citep{statista_tiktok_2025}. Through its “For You Page,” TikTok utilizes a recommendation algorithm that delivers personalized content streams based on user engagement~\citep{GRIFFITHS2024101807}. With more than 45.5 million TikTok posts associated with the hashtag \#GymTok as of May 2026 \citep{tiktok_gymtok}, users who demonstrate an interest in fitness may be presented with a virtually endless stream of body-related imagery, behaviors, and norms---raising concerns about the platform's potential implications for body image. Together, these features position TikTok as a uniquely influential environment for the dissemination and reinforcement of muscularity- and appearance-focused content.

Consequently, researchers have begun to examine the characteristics of fitness-related TikTok content and its effects on viewers' body image, finding that such content frequently promotes culturally idealized body types and objectifying portrayals~\citep{nuhn2025understanding}, which have been associated with increased body dissatisfaction via appearance comparison~\citep{pryde2022tiktok, de2025fittok, drivas2024whatieatinaday}. Notably, gender differences in content characteristics have been observed, with fitness TikTok videos featuring men more often emphasizing muscular ideals and face-obscuring forms of objectification, suggesting that male-oriented fitness content may present distinct appearance norms that remain underrepresented in existing literature \citep{pryde2024you}. The importance of identifying these unique norms is underscored by emerging experimental work, which found that exposure to idealized fitness and supplement TikTok content significantly decreased physical fitness satisfaction and increased intentions to use muscle-building substances in young men~\citep{beos2026impact}. These findings underscore the urgent need to examine how muscularity-oriented body ideals are uniquely conveyed and reinforced within the digital platforms most central to the lives of adolescent boys and young men.

Two complementary psychological frameworks help explain how exposure to body-focused content translates into body image outcomes. \textit{Social comparison theory} (SCT) posits that individuals evaluate themselves relative to others, particularly in domains such as physical appearance, where standards are subjective~\citep{festinger1954theory}. In highly visual social media environments, repeated exposure to idealized bodies may increase the opportunity for upward appearance comparisons and subsequent body dissatisfaction in men ~\citep{huang2021media, flynn2020relationship}.  Within masculinity-oriented contexts, such comparisons are also inherently tied to status and identity, as masculine status is often conceptualized as a socially negotiated and precarious construct, with physical appearance serving as a visible marker of social rank and masculine self-concept~\citep{vandello2013hard, pope2000adonis}. 

Complementing this perspective, \textit{objectification theory} suggests that individuals may internalize an external, appearance-focused perspective on the self, evaluating their bodies in terms of how they are seen by others~\citep{fredrickson1997objectification}. These processes are mutually reinforcing, forming cycles of comparison, evaluation, and self-surveillance that have been linked to muscularity-oriented concerns in men, such as social physique anxiety and muscle dysmorphia symptoms~\citep{brasil2024social,seekis2021social} and thinness-oriented concerns in women \citep{choukas2022perfect}. On platforms like TikTok, algorithmic curation may further intensify these dynamics by repeatedly reinforcing exposure to idealized bodies and normative standards, thereby embedding users in ongoing cycles of comparison and self-objectification~\citep{cotter2022fyp}. 

\subsection{Present Study}
Building on this framework, we investigate how muscularity-oriented content on TikTok both reflects and reinforces these dynamics. We analyze over 2,000 TikTok videos sourced from fitness-related keywords, annotated by clinical experts using a taxonomy of content themes (e.g., exercise, diet, masculinity, performance-enhancing substance use) and rated for their potential to harm viewers’ body image. We additionally characterize the muscularity of bodies depicted using a standardized scale  and examine how these characteristics relate to engagement metrics such as likes, views, and comments. These data allow us to explore the following research questions (RQs): 

\textbf{RQ1:} What are the dominant themes of \#GymTok content, and how are they  associated with visual representations of muscularity? 

\textbf{RQ2:} How does perceived harm to viewers’ body image vary across \#GymTok content themes, and to what extent does harm severity vary with the level of muscularity portrayed in these videos? 

\textbf{RQ3:} How do different \#GymTok content categories differ in user engagement, and to what extent is engagement associated with the level of muscularity depicted and the perceived harm of videos?

Our analysis reveals systematic trends in content prevalence, perceived harm, and levels of peer engagement. Muscularity self-objectification (e.g., videos featuring creators posing, flexing, or otherwise highlighting muscle definition) is the most prevalent theme, followed by videos encouraging strict dietary control and excessive exercise, reflecting a strong emphasis on appearance and behavioral regulation. Perceived harm varies markedly across categories, with videos about use of performance-enhancing substances rated as most harmful. Engagement also follows a distinct pattern: masculinity-focused content generates the highest levels of social validation, with most likes, views and comments. Notably, perceived harm and the muscularity of the bodies depicted in these videos are positively associated with engagement, such that more harmful content and more extreme physiques are rewarded within the platform’s attention economy.

Taken together, these findings point to systematic variation in both the production and algorithmic amplification of muscularity-oriented content on TikTok. Consistent with social comparison and objectification frameworks, the most engaging content is also the most harmful, potentially reinforcing idealized bodies and appearance-based norms. More broadly, these patterns suggest that TikTok may function as a digital environment in which status-driven competition is increasingly mediated through the body, with implications for the development of muscle dysmorphia and related pathologies.

\section{Methods}
\subsection{Taxonomy Development}
To guide video identification and classification, we employed a taxonomy of muscularity-oriented fitspiration content originally developed by \citet{chuBigTok} consisting of five primary categories: (1) Relationship to Body, (2) Relationship to Food, (3) Relationship to Exercise, (4) Supplement Abuse, and (5) Relationship to Masculinity. Each category includes clinically informed subcategories and operational definitions to support consistent identification and coding of relevant content. The full taxonomy can be found in the Supplemental Materials.

\subsection{Data Collection}
The video corpus for this study was sampled from a larger dataset compiled by \citet{chuBigTok} utilizing TikTok's Research API \citep{tiktok2026researchapi} to query videos posted between January 2019 and January 2025. This original corpus, \textit{BigTok}, was generated using 40 domain expert-curated query terms corresponding to taxonomy subcategories (e.g., ``swole,'' ``aggressive cut,'' ``tren''), with up to 1,000 videos queried per term. For each video, the total number of views, likes, comments, and shares was recorded at the time of data collection. From this corpus, 2,400 videos were randomly sampled for the current analysis; after removing duplicates and videos that could not be fully processed, the final analytic sample comprised 2,210 videos.

\subsection{Procedure}
Sixteen subject matter experts (13 female, 3 male), including clinical psychologists, social workers, and body image researchers, served as coders for the content analysis. This multidisciplinary expertise ensured that coding procedures were both theoretically informed and clinically grounded. Coders completed a structured training process using 20–30 example videos that were not included in the final dataset. Following training, coders met as a group to review coding procedures and resolve questions through discussion, ensuring consistent interpretation of the coding framework.

The dataset was divided into eight batches of approximately 300 videos. Each batch was independently coded by two annotators, such that every video received two independent ratings. For each video, coders evaluated the visual content, audio, and caption to identify the primary category and subcategory of harm. Coders could also assign a secondary category and subcategory when applicable. Coders also indicated whether the video included visible depictions of a person’s body and whether the content appeared to be commercially sponsored (yes/no). User identifiers were anonymized to strengthen privacy protections, and annotators were restricted to viewing only video content and captions.

In addition, coders rated the perceived potential for psychological harm to viewers using a 5-point Likert scale ranging from 1 (not harmful) to 5 (extremely harmful). Coders were trained to adopt the perspective of an impressionable young male viewer, a primary demographic for such content, who might encounter these videos via algorithmic recommendations on their social media feed. This interpretive lens required coders to assess harm as a function of how a developing audience might perceive the creator’s message and subsequently internalize the attitudes, behaviors, and aesthetic standards presented. Consequently, a high rating reflected the potential for the content to foster maladaptive social comparisons or to normalize high-risk behaviors, such as restrictive dieting or performance-enhancing drug use, by framing them as necessary components of masculine social validation.

\subsubsection{Intercoder Reliability}
To evaluate intercoder reliability for primary category and subcategory assignments, we calculated Cohen's $\kappa$ \citep{cohen1960coefficient}. Because the coding scheme is hierarchical, we assessed agreement between the two coders at two distinct levels of granularity: \textit{strong agreement} (both coders matched exactly on both the primary category and the subcategory) and \textit{weak agreement} (both coders matched on the primary category, but diverged on the specific subcategory). Following an initial independent coding phase, the coder pairs engaged in a structured consensus resolution process to discuss and reconcile contested edge cases. Post-consensus reliability was robust across all batches, with weak agreement (primary category) ranging from $\kappa = 0.78$ to $0.94$, and strong agreement (exact subcategory) ranging from $\kappa = 0.58$ to $0.81$.

For the ordinal measure of perceived harm severity (rated on a 1--5 Likert scale), we calculated the Intraclass Correlation Coefficient (ICC). The severity ratings demonstrated moderate agreement ($ICC = 0.413$), a level typical for subjective sociological and clinical assessments of harm. Notably, 69.4\% of all paired annotations between the two coders fell within one point of each other. The final severity score assigned to each video for downstream analysis was calculated as the mean of the two independent coders' ratings.

\subsection{Body Type Classification}
\label{sec:body_type}
\begin{figure}[ht!]
  \centering
      \includegraphics[width=0.9\linewidth]{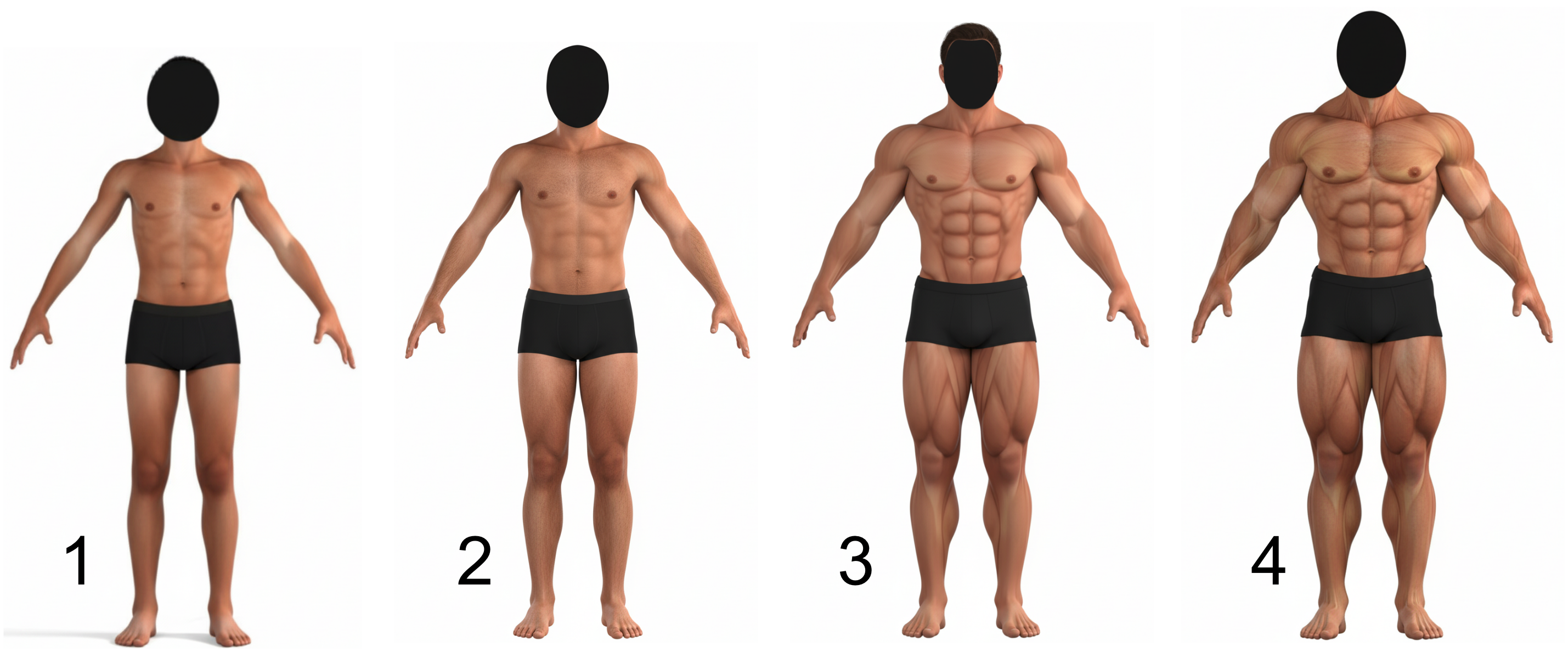}
  \caption{Body Type Classification. Visual Body Scale for Men is a muscularity figural rating scale, which classifies bodies into one of four categories: (1) underfat, (2) average, (3) athletic, and (4) bodybuilder~\protect\citep{talbot2019visual}.}
  \label{fig:body-type}
\end{figure}
To categorize the physiques of the individuals depicted in the videos, we employed an automated visual classification approach utilizing a Vision-Language Model Gemini 2.5 Flash \citep{comanici2025gemini25pushingfrontier}. Based on \citet{talbot2019visual}'s \textit{Visual Body Scale for Men}--Muscularity figural rating scale, visible bodies were classified into one of four target categories: underfat, average, athletic, and bodybuilder (see Figure~\ref{fig:body-type}). To ensure high classification accuracy, the model was implemented using a few-shot prompting strategy. Specifically, it was supplied with six anchor images representing these categories, alongside explicit textual descriptions of key visual traits for each body type. These anchor images were originally extracted from the \citet{talbot2019visual} study and subsequently computationally enhanced to improve fine-grained visual detail. This enhancement process was carefully calibrated to maintain the original quality and clinical authenticity of the scale, and the final reference images were unanimously approved by all authors before model deployment. During classification, the model was instructed to isolate the clearest full-body, frontal-view frames within each video and evaluate specific physiological markers—such as shoulder breadth, torso thickness, limb girth, and visible muscular definition. To prevent confounding variables, the model was constrained to ignore audio, captions, and background context. Bodies that were heavily obscured, out of frame, or otherwise unable to be categorized by the model were classified as ``unidentified.''

\section{Results}
We use the \#GymTok data to answer the research questions.

\subsection{Content Themes and Characteristics}
Our first research question asked: What are the most prevalent types of fitness-related content on TikTok, and what thematic and visual characteristics define them? To address this, we first report the distribution of content across categories and then describe the thematic and visual elements that characterize each group.
\begin{figure}[ht!]
  \centering
      \includegraphics[width=1\linewidth]{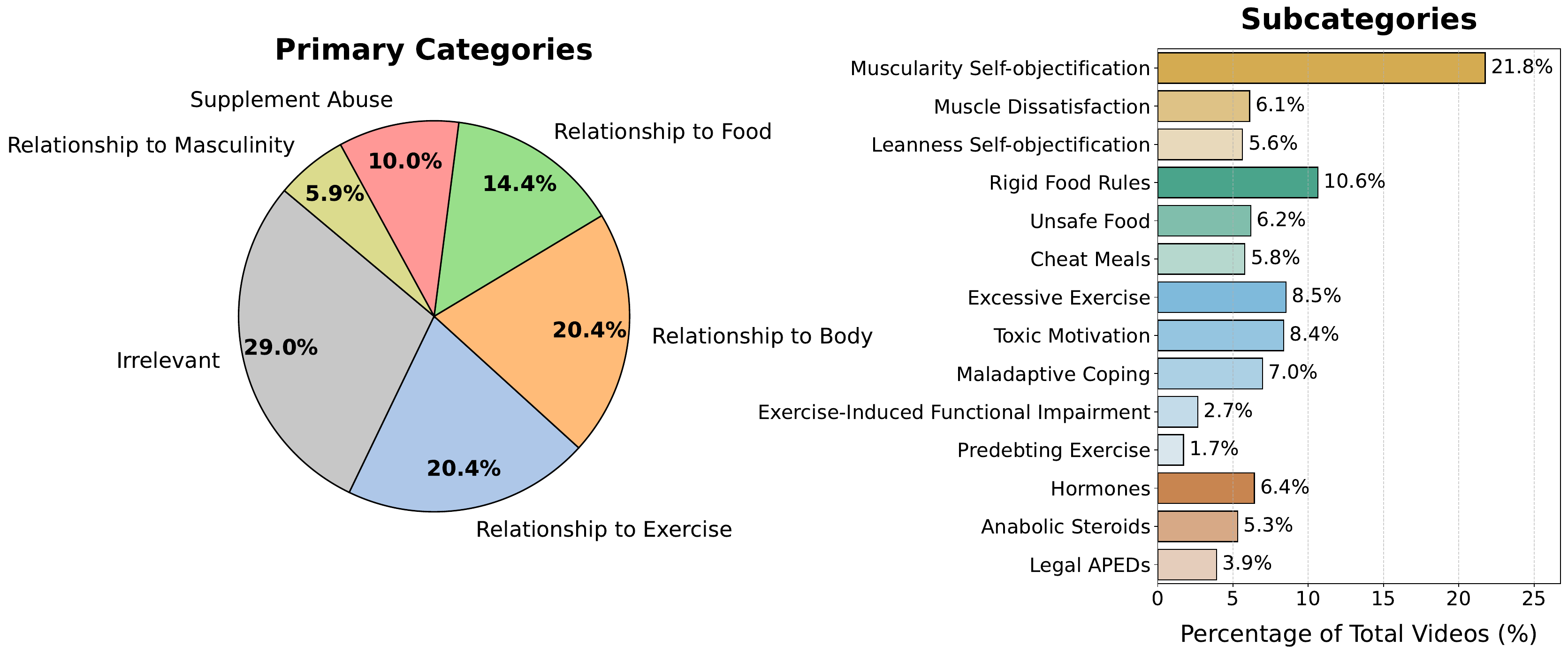}
  \caption{Distribution of videos classified by Primary Categories and Subcategories}
  \label{fig:cat-dist}
\end{figure}
\#GymTok content spanned a diverse set of thematic categories (Fig.~\ref{fig:cat-dist}). All percentages below are computed on the final analytic sample after excluding four videos rated as relevant but not harmful ($N = 2{,}206$); subcategory percentages are computed over the $n = 1{,}278$ videos assigned a specific subcategory. The largest shares of videos focused on relationships to exercise (20.4\%) and body (20.4\%), together accounting for roughly two-fifths of the sample. Food- and diet-related content comprised 14.4\% of videos, followed by videos related to supplements and steroids (10.0\%), and masculinity (5.9\%). Videos classified as irrelevant (29.0\%) did not contain substantive fitness, exercise, or body-related content (e.g., general food or lifestyle videos).

\subsubsection{Relationship to Exercise}
One of the two most frequently coded content categories was Relationship to Exercise, with 20.4\% of videos. Its most common subcategories were excessive exercise, toxic motivation, and maladaptive coping.

\textit{Excessive exercise} content accounted for 8.5\% of subcategorized videos ($n = 109$). These videos commonly depicted individuals training through visible signs of physical strain or exhaustion, such as trembling during repetitions, collapsing after sets, or continuing to lift despite pain. In these contexts, physical discomfort was framed not as a signal for rest, but as an expected---or even necessary---indicator of commitment and authenticity.

\textit{Toxic motivation} was present in 8.4\% of videos. These included point-of-view (POV) videos with motivational text overlays (e.g., ``POV: the weights are too heavy but you still train to failure'') as well as stylized montage edits featuring highly muscular figures accompanied by captions such as ``No days off.'' Collectively, these videos promoted perseverance and intensity, often without regard for physical or psychological limits.

Videos reflecting \textit{maladaptive coping} comprised 7.0\% of the sample. This content often emphasized the psychological benefits of exercise, frequently using hashtags such as ``\#gymtherapy.'' While some framed exercise as a tool for improving mental health, others portrayed it as a means of avoidance or emotional reliance. For example, one video featured a creator walking on a StairMaster with the caption: ``gym over everything because that’s the only thing that’s been there for me when I have nobody.''

\subsubsection{Relationship to Body}
Videos related to body image also comprised 20.4\% of the sample and were primarily characterized by self-objectification and muscle dissatisfaction. The most prevalent subcategory across the entire dataset was \textit{muscularity-oriented self-objectification} (21.8\%; $n = 278$). These videos framed exercise and dietary practices as means of achieving a muscular ideal, often featuring creators posing, flexing, or otherwise highlighting muscle definition and vascularity. In many cases, the body itself functioned as the central visual and narrative focus, with some videos consisting almost entirely of physique monitoring (i.e. “body checking”), featuring repeated posing to assess muscular development, vascularity, and leanness.

A smaller subset (6.1\%) was categorized as \textit{muscle dissatisfaction}. In these videos, content creators explicitly expressed dissatisfaction with their physique, often through visual self-comparison paired with captions such as ``not big enough.'' Some also framed muscularity as an ongoing, never-complete process, including transformation narratives and progress tracking.

Finally, 5.6\% of videos depicted \textit{leanness-oriented self-objectification}, emphasizing low body fat and visible abdominal definition. These videos frequently included physique checks, workout routines, or dietary content framed around achieving or maintaining a lean appearance.

\subsubsection{Relationship to Food}
Approximately 14.4\% of videos were primarily about food and nutrition. The largest subset involved \textit{rigid food rules} (10.6\%), often presented through ``full day of eating'' videos or meal-prep content accompanied by macronutrient tracking and dietary restrictions. Many of these videos focused on weight loss through ``cutting,'' sometimes emphasizing extreme approaches such as ``mini cuts'' or ``aggressive cuts,'' with caloric intake as low as 1,200–1,400 calories per day and rapid weight loss over short periods. These videos frequently featured before-and-after comparisons, with captions emphasizing discipline and strict adherence (e.g., ``There was only one day where I ate over those calories''), framing rigidity as both necessary and admirable.

\textit{Unsafe food} consumption accounted for 6.2\% of subcategorized videos, primarily involving raw animal products (e.g., raw meat, animal organs, or eggs). These videos often included claims about enhanced nutritional benefits or protein absorption, with little or no mention of potential health risks. Reflecting a viral trend within GymTok from 2023, a few videos included creators eating dog food ~\citep{buzzfeednewsBrosEating}. This phenomenon was largely fueled by misleading data within nutritional tracking apps, which erroneously suggested that certain pet foods contained extreme protein densities, as well as creators leveraging the shock value of eating unpalatable food to achieve virality within the GymTok community.

\textit{Cheat meals} accounted for 5.8\% of the sample. These videos showcased high-calorie meals framed as deviations from otherwise restrictive diets, often positioned as indulgent or contradictory to fitness goals—messaging that has been linked to disordered eating patterns~\citep{tsang2025role}. Relatedly, some videos depicted exercise as a means to ``earn'' food intake (e.g., ``training so you can eat more food''), reinforcing transactional relationships between food and physical activity.

\subsubsection{Supplement Abuse}
Approximately 10.0\% of videos featured performance-enhancing supplements or steroids. This included \textit{hormone replacement therapies} (6.4\%), \textit{anabolic steroids} (5.3\%), and \textit{legal appearance- and performance-enhancing drugs} (APEDs; e.g., creatine, protein powder; 3.9\%). Content ranged from promotional videos highlighting benefits---often supported by anecdotal or purported scientific evidence---to cautionary content describing adverse effects such as acne, infertility, or cardiovascular risks. Some videos also employed humor or satire to depict APED use, including exaggerated or dramatized portrayals.

\subsubsection{Relationship to Masculinity}
The smallest category (5.9\%) focused explicitly on masculinity. These videos linked muscularity to broader ideals of heteronormative masculinity, positioning a highly muscular body as a marker of strength, discipline, and social dominance. Unlike other categories, this content more frequently relied on curated compilations of athletes, fictional characters, or cultural icons, accompanied by motivational audio emphasizing resilience, self-mastery, and the rejection of weakness.

\subsubsection{Sponsorship}
Videos were further coded for the presence or absence of sponsorship. Approximately 5.7\% ($n = 119$) of the videos were tagged as having sponsored content. The Supplement Abuse category featured the most sponsored videos, with 22.2\% of videos including sponsored content. The Relationship to Masculinity category had the lowest amount of sponsored content, with 0.8\% of videos featuring commercial content.

\begin{figure}[ht!]
  \centering
      \includegraphics[width=0.9\linewidth]{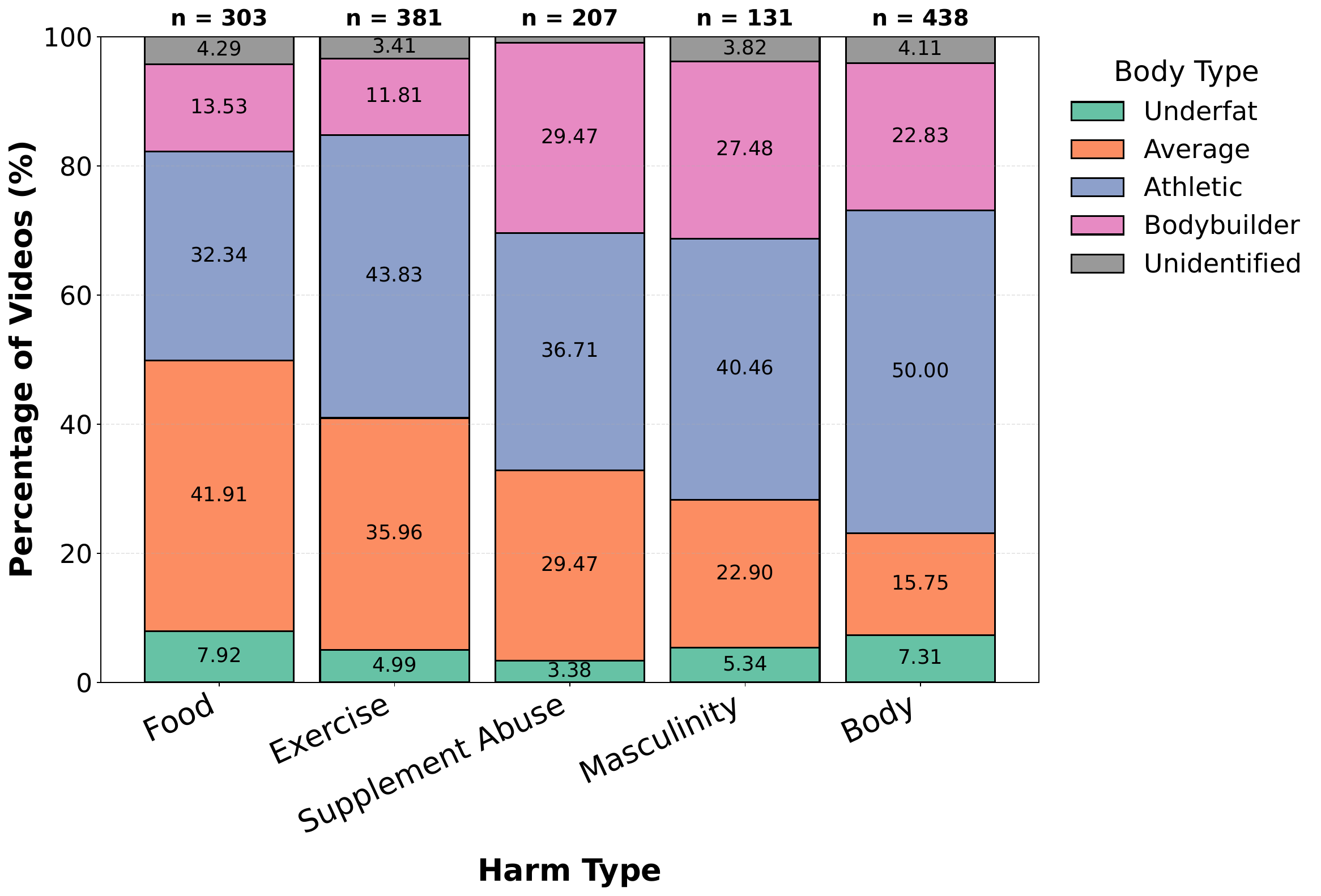}
  \caption{Distribution of Body Types by Primary Content Category}
  \label{fig:body-type-distribution}
\end{figure}
\subsubsection{Body Type}
Videos were coded for the presence or absence of a visible body. Across the dataset, a visible body was present in 42.4\% of videos ($n = 893$). Body visibility was highest in the Relationship to Body category (89.8\%) and, among the harm-relevant categories, lowest in the Supplement Abuse category (43.0\%).

Among videos with visible bodies, muscularity was coded using a standardized scale (Underfat, Average, Athletic, Bodybuilder; see Section~\ref{sec:body_type}). Body type distributions varied by category (Fig.~\ref{fig:body-type-distribution}). Food-related content displayed the greatest diversity, with the highest proportion of Average physiques. Exercise-related videos contained relatively few Bodybuilder physiques, instead featuring a mix of Athletic and Average bodies. In contrast, body image content showcased muscular bodies, with the majority of videos depicting Athletic or Bodybuilder physiques. Supplement Abuse and Masculinity-related content also featured more frequent representations of Bodybuilder physiques, with Underfat bodies being least represented in the Supplement Abuse category. 

\begin{figure}[ht!]
  \centering
      \includegraphics[width=0.7\linewidth]{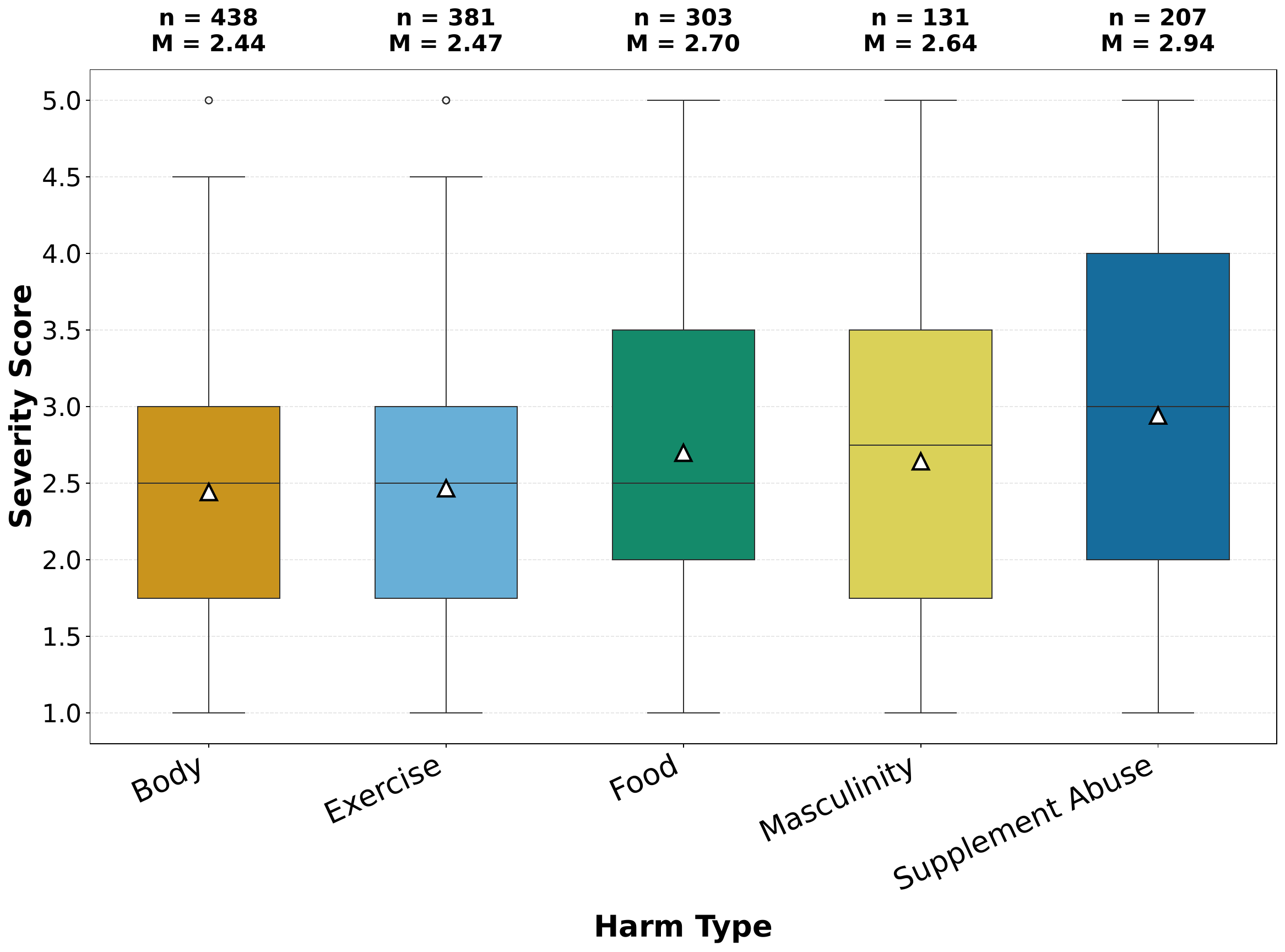}
  \caption{Distribution of perceived harm ratings across content categories. Ratings reflect expert assessments of the potential for video to negatively influence viewers’ body image, measured on a 5-point scale from 1 (not harmful) to 5 (extremely harmful). Boxplots show the median (center line),  mean (triangle), interquartile range (25th–75th percentiles; box), and observations within 1.5 times the interquartile range (whiskers); points beyond whiskers represent outliers.}
  \label{fig:harm-severity}
\end{figure}

\subsection{Perceived Harm to Body Image}
\label{sec:harm}
The second aim of this study was to examine how different types of muscularity-oriented content vary in their potential to negatively impact viewers’ body image. Perceived harm was operationalized as coders’ ratings on a 5-point Likert scale, ranging from 1 (not harmful) to 5 (extremely harmful).

Perceived harm ratings spanned the full range of the scale, with most videos clustered in the low-to-moderate range ($N = 2{,}099$; $M = 2.12$, $SD = 1.09$). The distribution was right-skewed, with relatively few videos receiving the highest severity ratings. 

To account for this imbalance and facilitate clearer comparisons across content types, ratings were grouped into three ordinal harm categories: Low ($\leq 1.5$), Medium ($> 1.5$ and $\leq 3.5$), and High ($> 3.5$). This categorization enables comparisons across substantively distinct levels of perceived harm while preserving the underlying ordinal structure of the data.

The Low harm category ($n = 932$) comprises videos that are primarily neutral or observational in nature. Although such videos may depict behaviors associated with body image risk, such as body checking or exercise framed around aesthetics, they do not explicitly promote or endorse these behaviors. For example, a typical video might feature a muscular individual flexing in a mirror with background music and a caption such as ``\#flexfriday,'' presenting the body as an object of display without overt prescriptive messaging.

In contrast, the Medium harm category ($n = 942$) captures more nuanced content. These videos often depict potentially harmful behaviors (e.g., restrictive dieting, intense training routines) but include mitigating elements such as disclaimers, personal framing, or humor. Some creators explicitly acknowledge the extremity of their habits or present them in a self-aware or satirical manner. While such content may still prompt upward comparison or normalize risky practices, these framing devices introduce a degree of ambiguity that distinguishes it from unqualified endorsement.

The High harm category ($n = 225$) is characterized by explicitly prescriptive or promotional content. Videos in this group actively advocate for behaviors such as steroid use, extreme dieting, eating unsafe food, or compulsive exercise, framing them as necessary or desirable for achieving an ideal physique. At this level, content shifts from personal experience to the promotion of normative standards, often positioning harmful behaviors as requirements for success. For instance, creators may depict steroid use as a form of commitment to fitness and directly encourage viewers to adopt similar practices.

\subsubsection{Harm Severity by Content Category}
Harm severity varied systematically across content categories (Fig.~\ref{fig:harm-severity}; full descriptive statistics are reported in Table~\ref{tab:severity_type_descriptives}, \ref{app:stats}). Masculinity-focused and supplement- and steroid-related content was rated more harmful on average, with median ratings approaching the high-harm threshold ($> 3.5$) and distributions extending into the highest severity levels. Food-related content was rated as moderately severe, with medians in the midrange but substantial variability. Body- and exercise-related content was rated least harmful on average ($M$s = 2.44 and 2.47, respectively), with larger proportions of videos clustered near the low-harm threshold ($\leq 1.5$), although some videos in these categories were also rated as highly harmful.

Post-hoc pairwise comparisons using Dunn’s test (Table~\ref{tab:severity_type_dunn}, \ref{app:stats}) further clarified these differences. Supplement-related content was rated as significantly more harmful than both body- and exercise-related content (\textit{p}s < .001), while food-related content was rated as more harmful than body-related content (\textit{p} = .016), with the corresponding contrast against exercise-related content falling just short of significance (\textit{p} = .052). No significant differences emerged among supplement, masculinity, and food categories, nor between body and exercise categories. These results suggest two broad clusters of content: lower-severity categories (Relationship to Exercise, Relationship to Body) and higher-severity categories (Relationship to Masculinity, Supplement Use, Relationship to Food), rather than fully distinct differences across all groups.

\begin{figure}[ht!]
  \centering
      \includegraphics[width=0.7\linewidth]{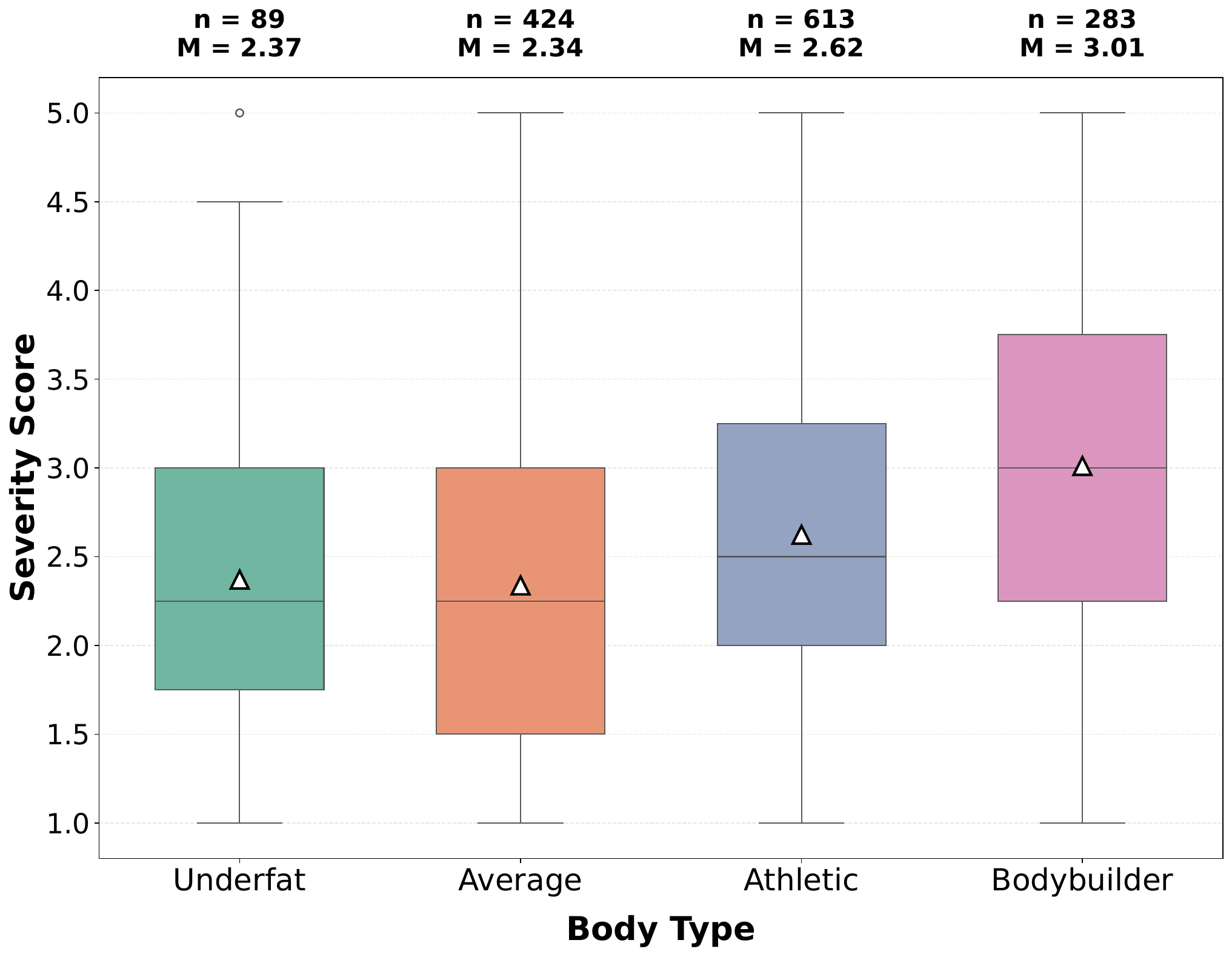}
  \caption{Distribution of perceived harm ratings by depicted body type. Body types were classified using the muscularity figural rating scale (Section~\ref{sec:body_type}); ratings reflect expert assessments of the potential for a video to negatively influence viewers’ body image, measured on a 5-point scale from 1 (not harmful) to 5 (extremely harmful). Boxplots show the median (center line), mean (triangle), interquartile range (25th–75th percentiles; box), and observations within 1.5 times the interquartile range (whiskers); group sizes ($n$) and means ($M$) are displayed above each distribution.}
  \label{fig:harm-muscularity}
\end{figure}

\subsubsection{Harm Severity by Muscularity}
Perceived harm also varied systematically with the muscularity of the bodies depicted (Fig.~\ref{fig:harm-muscularity}; $n = 1{,}409$ harm-relevant videos with a codable body type). Mean harm ratings increased monotonically across the muscularity scale, from average ($M = 2.34$) and underfat ($M = 2.37$) physiques to athletic ($M = 2.62$) and bodybuilder ($M = 3.01$) physiques (Kruskal--Wallis $H(3) = 84.93$, \textit{p} < .001). Videos featuring bodybuilder physiques were rated as most harmful, with median ratings of 3.0 and distributions extending well into the high-harm range, whereas videos depicting underfat and average bodies clustered toward the lower end of the scale.

Post-hoc pairwise comparisons using Dunn’s test (Table~\ref{tab:severity_bodytype_dunn}, \ref{app:stats}) indicated that bodybuilder content was rated as significantly more harmful than every other body type (\textit{p}s < .001), and athletic content was rated as more harmful than average content (\textit{p} < .001). Ratings did not differ between underfat and average physiques (\textit{p} = 1.00), and the contrast between underfat and athletic physiques fell short of significance (\textit{p} = .069). Together, these results indicate that perceived harm scales with the muscularity of the bodies displayed: content showcasing the most extreme physiques was also judged most likely to negatively influence viewers’ body image.

\subsection{Peer Engagement with \#GymTok Videos}
\label{sec:engagement}
The third research question examined how content characteristics and perceived harm of \#GymTok content relate to engagement. Engagement metrics capture the extent to which users view, like, share, and comment on videos, serving as indicators of both audience attention and interaction. Importantly, these signals are also used by TikTok’s recommendation system to rank and distribute content to broader audiences. Because videos that generate higher engagement are more likely to be surfaced in users’ feeds, engagement functions not only as a measure of popularity but also as a mechanism through which content and social comparison and objectification processes are algorithmically amplified. We compared engagement metrics across three dimensions: primary content category (excluding videos classified as ``Irrelevant''), depicted body type, and perceived harm severity.

\begin{figure}[ht!]
  \centering
    \includegraphics[width=0.95\linewidth]{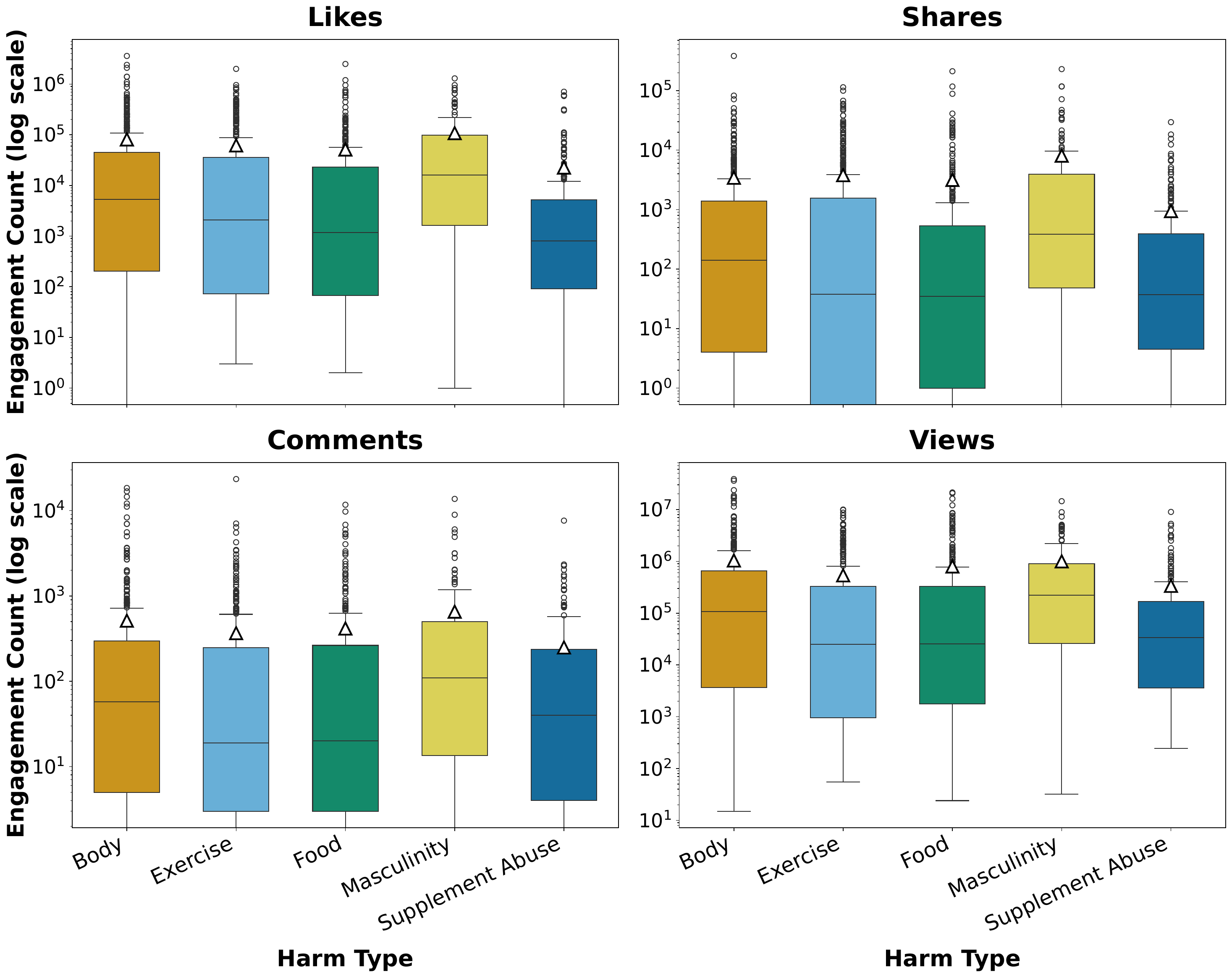}
  \caption{Engagement Metrics by Primary Content Category. To account for the highly right-skewed distribution of engagement data, the y-axis is plotted on a logarithmic scale. Boxplots show the median (center line),  mean (triangle), interquartile range (25th–75th percentiles; box), and observations within 1.5 times the interquartile range (whiskers); points beyond whiskers represent outliers. Content categories are ordered based on their median perceived harm severity (Section \ref{sec:harm}.)}
  \label{fig:engagement-category}
\end{figure}

\subsubsection{Engagement by Content Category}
Engagement patterns varied substantially across \#GymTok content categories (Fig.~\ref{fig:engagement-category}). Masculinity-focused videos received the most engagement, recording the highest median number of likes, shares, views, and comments, and the highest mean engagement on all metrics except views, for which body-related content was comparable. In contrast, supplement-related content showed the lowest mean engagement across all metrics and the lowest median number of likes, although its median shares, comments, and views were comparable to those of exercise- and food-related content. Full pairwise statistics for these category differences are reported in Table~\ref{tab:engagement_type_dunn} (\ref{app:stats}).

Across all metrics, engagement distributions were heavily right-skewed, with mean values substantially exceeding medians. This pattern indicates that overall engagement is driven by a small number of highly viral videos.

\begin{figure}[ht!]
  \centering
    \includegraphics[width=0.8\linewidth]{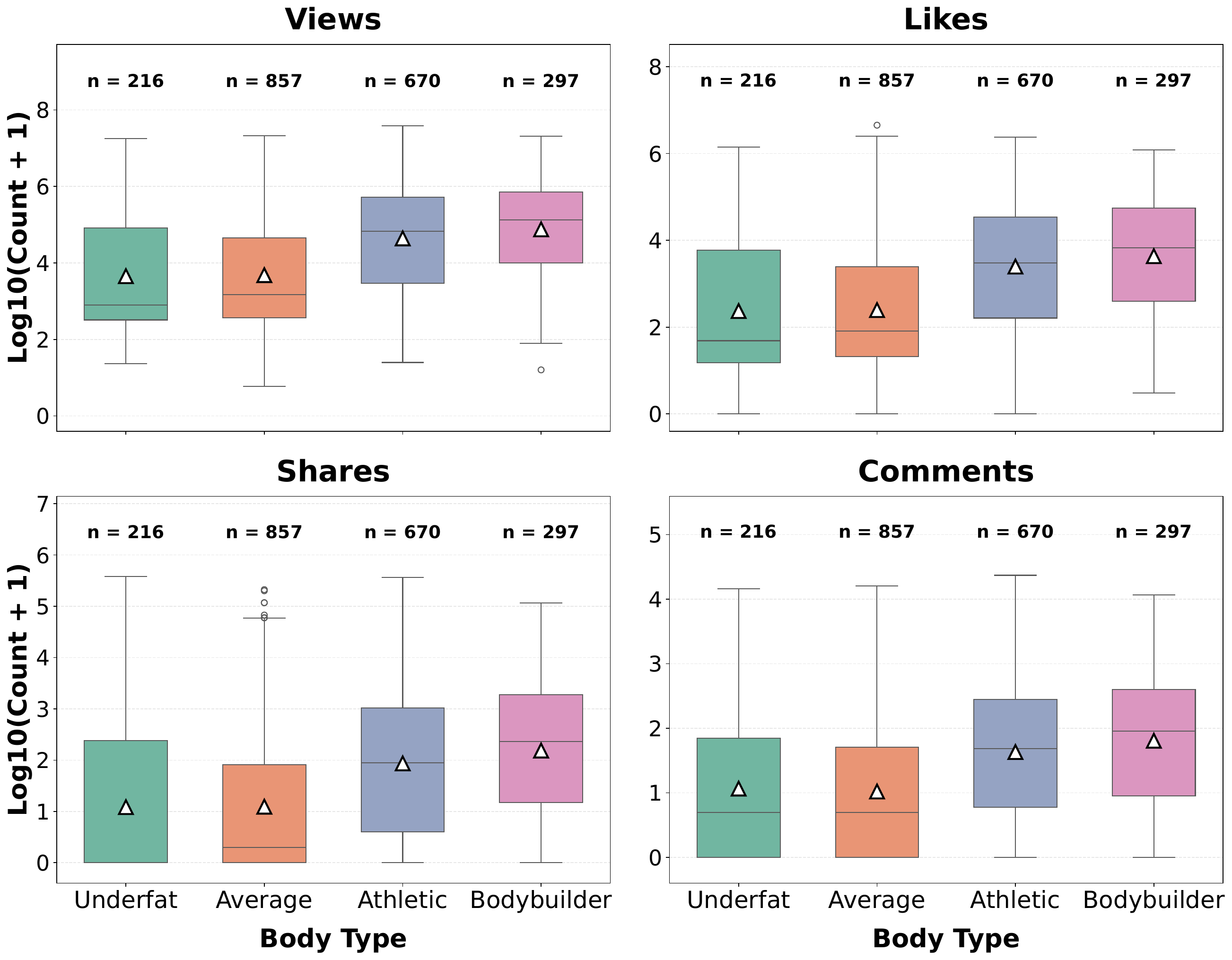}
  \caption{Engagement Metrics by Body Type. To account for highly right-skewed distributions, raw engagement counts were mathematically log-transformed ($\log_{10}(x+1)$) prior to plotting. White triangles within the boxes denote the mean value for each group, and specific sample sizes ($n$) are displayed above each distribution. Categories along the x-axis are ordered by increasing levels of visible muscularity.}
  \label{fig:engagement-bodytype}
\end{figure}
\subsubsection{Engagement by Body Type}
Engagement also varied systematically by depicted body type (Fig.~\ref{fig:engagement-bodytype}). Videos featuring athletic and bodybuilder physiques received higher engagement across all metrics. In contrast, videos featuring underfat and average bodies showed consistently lower central tendencies, although all groups exhibited substantial variability.

As with content categories, engagement distributions were heavily skewed, with long upper tails indicating that a small proportion of videos achieved high virality across all body types. Nevertheless, the overall pattern suggests that \textit{more muscular physiques are associated with greater audience engagement} in terms of likes, shares, comments, and views.

Pairwise comparisons further clarified these differences (Table~\ref{tab:engagement_bodytype_mwu}, \ref{app:stats}). No statistically significant differences were observed between underfat and average body types across any engagement metric (\textit{p}s = .771–1.00). In contrast, comparisons between the lower-muscularity groups (underfat, average) and higher-muscularity groups (athletic, bodybuilder) were consistently significant across all metrics (all \textit{p}s $< .001$), indicating a clear separation in engagement between these clusters.

Within the higher-muscularity categories, differences between athletic and bodybuilder physiques were largely nonsignificant for views (\textit{p} = .073), likes (\textit{p} = .064), and comments (\textit{p} = .066), suggesting similar engagement patterns. A modest but statistically significant difference emerged for shares (\textit{p} = .032), indicating slight variation in this specific form of engagement. Overall, these findings point to two primary clusters—lower-muscularity (underfat, average) and higher-muscularity (athletic, bodybuilder)—with substantial differences between clusters but minimal differentiation within them.

\begin{figure}[ht!]
  \centering
    \includegraphics[width=0.75\linewidth]{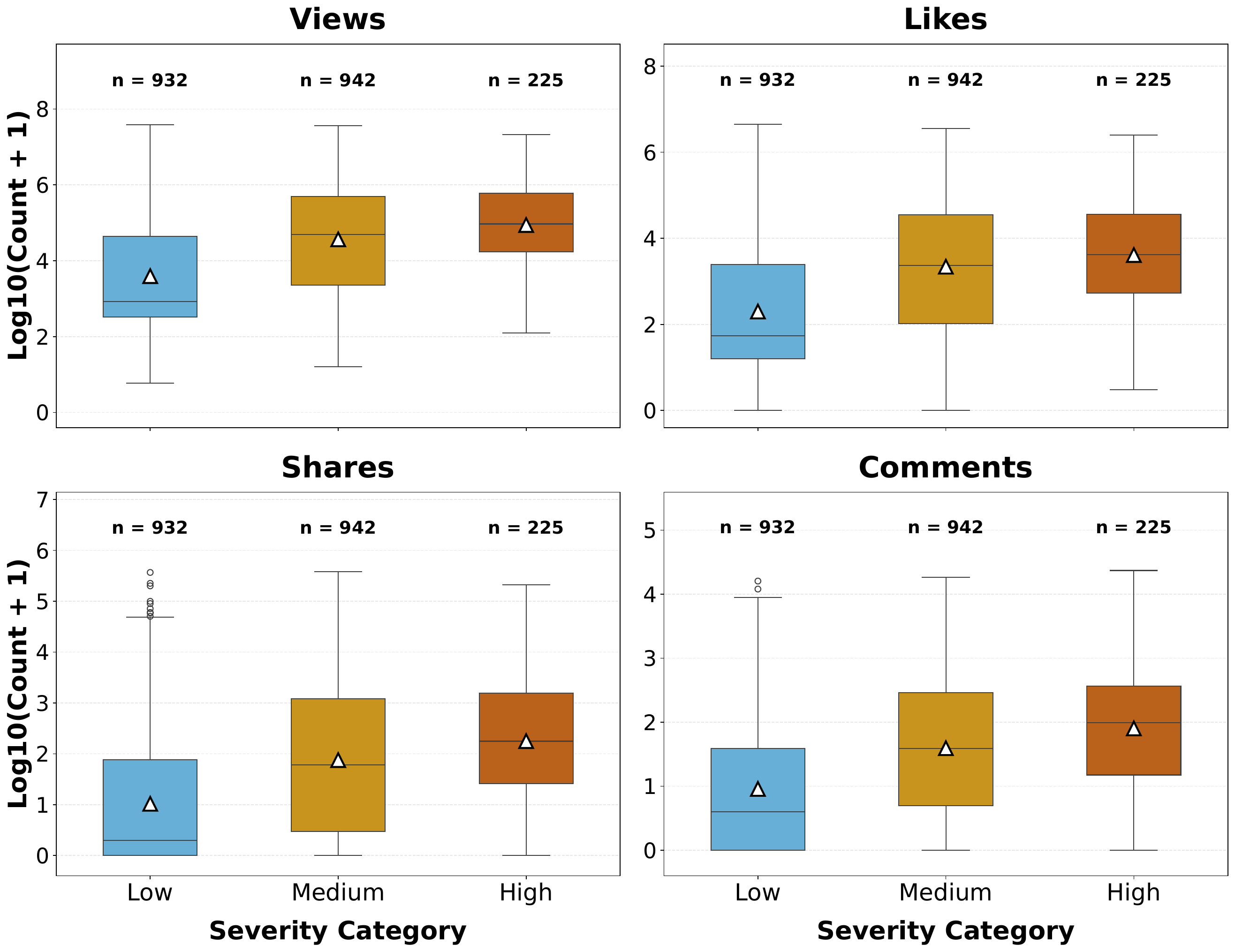}
  \caption{Engagement Metrics by Perceived Harm Severity Category. Videos were categorized into Low ($\le 1.5$), Medium ($1.5 < x \le 3.5$), and High ($> 3.5$) severity groups based on their perceived harm ratings. To account for highly right-skewed distributions, raw engagement counts were mathematically log-transformed ($\log_{10}(x+1)$) before plotting. White triangles within the boxes denote the group means, and specific sample sizes ($n$) are displayed above each distribution.}
  \label{fig:engagement-harm}
\end{figure}

\subsubsection{Engagement by Harm Severity}
To examine differences in engagement across levels of perceived harm to viewer's body image, harm ratings were categorized as low ($\leq 1.5$), medium ($> 1.5$ and $\leq 3.5$), and high ($> 3.5$). Engagement increased monotonically across these categories, with \textit{high-severity videos receiving the greatest engagement}, followed by medium-severity videos (Fig.~\ref{fig:engagement-harm}).

Post hoc pairwise comparisons (Table~\ref{tab:engagement_severity_dunn}, \ref{app:stats}) indicated significant differences between all severity groups, demonstrating a consistent increase in engagement from low to medium to high severity across all metrics (all \textit{p}s $\leq .036$).

\section{Discussion}
Taken together, our findings suggest that \#GymTok is not merely a repository of fitness videos, but a digital environment in which muscularity functions as a highly visible status signal, and content depicting muscular physiques is systematically rewarded through user engagement and algorithmic amplification---even when that content promotes potentially harmful ideals and behaviors. As a result, the platform may reinforce appearance-based standards that heighten body image risk among boys and young men. In what follows, we discuss our main findings through four key insights.

\subsection{\#GymTok as Status Competition in Modern Masculinity}
A central finding of this study is that masculinity-focused content received the highest levels of online engagement, despite constituting a relatively small portion of the overall sample. This amplification suggests that \#GymTok is not simply concerned with fitness or aesthetics, but also with the performance and negotiation of masculine status. Consistent with theories of status-driven risk taking and the “young male syndrome'' \citep{wilson1985competitiveness,ashton2010status}, young men have long engaged in competitive behaviors to signal dominance, competence, and social value. Our findings indicate that these dynamics persist in digital environments, where status is publicly displayed and quantified through engagement metrics.

One possible explanation for the outsized popularity of masculinity-oriented content is its close alignment with cultural norms surrounding male identity, strength, and status, rendering it particularly resonant for young male audiences. Prior research suggests that fitspiration content shapes men's attitudes toward masculinity and encourages comparisons not only of physique but also of identity and personal worth \citep{palmer2015a}. Within \#GymTok, muscularity appears to function as a form of identity capital---a visible resource through which individuals construct, perform, and communicate valued aspects of masculinity and social competence~\citep{cranswick2025s}. By framing muscularity as a marker of masculine worth rather than merely an aesthetic ideal, such content may engage viewers not only at the level of appearance but also at the level of identity and belonging. From a social identity perspective, the body becomes a visible signal of group membership, encouraging viewers to evaluate themselves against normative masculine ideals rather than appearance standards alone.

These dynamics may be especially salient for adolescent boys and young men, who are actively negotiating masculine identity and may be particularly sensitive to identity-relevant appearance standards and peer evaluation \citep{choukas2022perfect}. In this context, the drive for belonging may supersede health concerns, leading young men to perceive high-risk behaviors, such as performance-enhancing drug use or extreme dietary restriction, not as dangers, but as badges of membership and discipline~\citet{nourse2024masculine}. The convergence of identity salience, peer validation, and algorithmic visibility thus positions \#GymTok as an environment in which status competition is not merely reflected but actively structured by platform dynamics.

\subsection{Muscularity as a Status Signal}
Beyond thematic differences, our results demonstrate that muscularity itself is systematically rewarded on \#GymTok. Videos featuring more muscular physiques received significantly higher engagement than those depicting less muscular bodies. While the overall prevalence of muscular bodies within fitness-related content is consistent with established analyses of TikTok \citep{nuhn2025understanding, pryde2024you} and Instagram \citep{angrish2024flex, gultzow2020male, deighton2018objectifying}, our findings extend this literature by demonstrating that these physiques are not distributed evenly across content themes. The most extreme muscular ideals were concentrated within supplement abuse and masculinity-related content, suggesting that the most prominent visual status signals co-occur with higher-risk behavioral norms.

Conversely, diet-related videos featured the greatest diversity of body types, suggesting that everyday health behaviors are presented as more accessible to the average user. This divergence is noteworthy: within \#GymTok, "attainable" health appears to be decoupled from ideal masculinity, potentially signaling to young men that achieving peak social status requires moving beyond conventional wellness into high-risk territory. The body type landscape of \#GymTok thus reinforces a hierarchy in which muscularity is not only more visible but more socially valued, while ordinary physiques are confined to lower-status content domains.

These patterns are further reinforced by platform architecture. On social platforms such as TikTok, engagement metrics play a central role in content ranking and recommendation, with algorithms leveraging signals such as views, likes, shares, and comments to determine which videos are surfaced to broader audiences \citep{tiktok2020foryou}. As a result, content that attracts higher engagement is systematically amplified, increasing its visibility and likelihood of repeated exposure. Consequently, videos depicting highly muscular physiques are more likely to be promoted, reinforcing the salience of muscular ideals and their association with status and desirability. The combination of visual salience and algorithmic reinforcement positions muscular bodies as dominant reference points, shaping the standards against which users evaluate themselves and potentially narrowing the range of body types encountered during routine platform use.

\subsection{Algorithmic Amplification of Harmful Content}
A particularly striking finding is the positive association between perceived harm and engagement. Videos rated as more harmful by expert annotators consistently received higher levels of engagement across all metrics, suggesting that the content most likely to promote risky or maladaptive behaviors is also the most visible and widely disseminated. This relationship reflects not only user preferences but also platform dynamics. By prioritizing highly engaging content, TikTok's recommendation algorithm may amplify harmful content, increasing its reach and frequency of exposure. In this way, the platform may inadvertently reinforce and normalize behaviors that carry elevated risk for body image disturbance.

This pattern is most acutely demonstrated in APED-related content. Although these videos comprised only a small proportion of the sample (10.0\%), they were rated as having the highest potential for harm, despite the display or promotion of such products constituting a violation of TikTok's community guidelines \citep{tiktok2024_mental_behavioral_health}. Critically, approximately 78\% of these videos were generated by non-sponsored users, suggesting that APED discourse is normalized not through commercial promotion but through peer-to-peer discussion. When references to performance-enhancing substances appear in non-sponsored content from ordinary users, they may be perceived less as high-risk medical decisions and more as accessible, peer-validated tools for achieving the muscular ideal \citep{nourse2024masculine, courtenay2000constructions}. The Tripartite Influence Model \citep{thompson1999theory} underscores the potency of precisely this kind of peer and media influence in shaping the internalization of body ideals and engagement in muscle-building behaviors, including steroid use \citep{ganson2025associations, tylka2011refinement}.

Importantly, however, risk within \#GymTok extends beyond explicit substance use. Masculinity-related content was also rated among the most harmful categories, indicating that harm is embedded in identity-level messaging that links muscularity to masculine worth. This convergence of high engagement and high harm across both behavioral and identity-oriented content suggests a structural alignment between what is rewarded by the platform and what is risky for viewers.

\subsection{Social Comparison, Self-Objectification, and Risk Amplification}
These patterns can be understood through the combined lenses of social comparison and objectification theories. \#GymTok provides a highly visual environment in which users are repeatedly exposed to idealized bodies alongside engagement signals that indicate which appearances are most valued. This creates fertile conditions for upward social comparison \citep{festinger1954theory}, particularly when highly muscular physiques are consistently associated with greater visibility and approval. At the same time, the prevalence of self-objectifying content alongside rigid dietary practices, excessive exercise, and physique-focused presentation encourages users to evaluate their own bodies from an external, appearance-based perspective \citep{fredrickson1997objectification}. Prior work has established that exposure to idealized muscular imagery promotes both social comparison and self-objectification among young men \citep{fatt2019effects, daniel2014objectification}, and our findings suggest that \#GymTok may intensify these processes by coupling idealized content with visible markers of social approval.

Crucially, these mechanisms are mutually reinforcing. That is, user engagement amplifies exposure to idealized bodies, which intensifies comparison, which in turn promotes further self-objectification and continued engagement with appearance-focused content. The link between harm and engagement suggests that this cycle is systematically intensified by the platform's reward structure, with \textit{content that most strongly encourages comparison and self-surveillance also the most likely to be algorithmically amplified}. Within this loop, behaviors such as APED use, rigid dieting, and excessive exercise function not as aberrant practices but as status-driven risk-taking strategies; a willingness to incur physical and psychological costs in pursuit of socially rewarded outcomes, namely muscularity, visibility, and belonging. These findings point to a self-reinforcing feedback loop in which the platform does not merely reflect user preferences but actively shapes them, positioning \#GymTok as a system of algorithmically mediated status competition in which the content carrying the greatest risk is structurally positioned for the widest reach.

\subsection{Risky Behaviors as Pathways to Pathology}
The self-reinforcing dynamics described above carry implications not only for normative body dissatisfaction but also for the development of clinical pathology. Within \#GymTok, behaviors such as APED use, rigid dieting, and excessive exercise are not only present but embedded within highly engaging content, often framed as necessary, aspirational, or indicative of discipline. When these practices are modeled by peers, validated through engagement, and algorithmically amplified, they may contribute to the escalation of subclinical body image concerns into clinical conditions, including muscle dysmorphia and disordered eating \citep{hilkens2021social, ganson2025associations}. In this sense, \#GymTok may function as a context in which aspirational self-improvement shifts toward pathological self-regulation.

This trajectory can be situated within a broader understanding of male risk-taking behavior. From an evolutionary perspective, young men's willingness to accept physical and social costs in pursuit of status is well documented~\citep{wilson1985competitiveness, ashton2010status}, and our findings suggest that digital environments may provide novel arenas for these tendencies to manifest. Within \#GymTok, the costs take specific forms, such as disordered eating, overtraining, and substance use, while the rewards are quantified through likes, shares, and followers. The platform thus transforms what might otherwise remain private health-risk behaviors into publicly performed and socially reinforced status displays. Importantly, however, not all young men who encounter this content will engage in these behaviors, raising the question of what individual and contextual factors moderate the pathway from exposure to risk. Future research should examine the role of psychological factors such as muscular ideal internalization, contingent self-worth, and body dissatisfaction \citep{tylka2011refinement}, as well as social variables including peer group norms, perceptions of the prevalence of risk behaviors, and the centrality of masculine identity to one's self-concept in considering the potential effects of engagement with this content \citep{choukas2022perfect, nourse2024masculine}.

\section{Conclusion}
This study analyzed a sample of 2,210 muscularity-oriented fitspiration TikTok videos (i.e., \#GymTok), with the goal of understanding the landscape of this content. Our findings show that \#GymTok frequently depicts highly muscular bodies, rigid exercise and dietary practices, APED use, and masculinity-linked messaging, collectively reinforcing narrow and appearance-focused standards of the male body. Content perceived as more harmful by clinical experts received significantly greater engagement, indicating that engagement-driven platform dynamics may disproportionately amplify extreme and potentially harmful representations of muscularity and masculinity. In this way, the platform not only reflects existing ideals but may also contribute to their reinforcement and normalization through algorithmic visibility. Given the well-documented effects of fitspiration and muscular-ideal content on image-based social media platforms, these findings highlight the importance of understanding how algorithmically curated platforms may influence the production, amplification, and social valuation of body ideals among boys and young men.

\section*{Ethics statement}
This study involved a secondary analysis of publicly available TikTok videos retrieved through the TikTok Research API. The research did not involve any interaction or intervention with human participants. All user identifiers were removed prior to coding, and annotators viewed only video content and captions. The study therefore did not meet the definition of human subjects research requiring institutional review board approval.

\section*{Data availability}
The videos analyzed in this study were obtained through the TikTok Research API and are subject to TikTok's Terms of Service, which restrict public redistribution of the raw data. The coding taxonomy, annotation scheme, and aggregated data supporting the findings of this study are available from the corresponding author upon reasonable request.

\section*{Declaration of generative AI and AI-assisted technologies in the manuscript preparation process}
During the preparation of this work, the authors used Claude (Anthropic) for improving the language and readability of the manuscript. The authors reviewed and edited the output as needed and take full responsibility for the content of the published article.

\section*{Acknowledgements}
We express our sincere gratitude to our collaborators, Roxanna Sharif, Ross Sonnenblick, Laura D'Adamo, Krista Tabuenca, Artha Gillis, Colleen McCord, Brandy Saccacio, Amanda Velkova, Chessa Kabiling, Jenny Nguyen, and Danielle Crowe, for generously volunteering their time and expertise to annotate the sensitive content. This work would not have been possible without their invaluable insights.

\section*{Funding}
This research did not receive any specific grant from funding agencies in the public, commercial, or not-for-profit sectors.

\section*{Declaration of Interest}
Stuart B. Murray and Kristina Lerman have offered expert witness testimony on matters relating to the topic of this paper. The remaining authors have no declaration of interest.

\section*{Author Contributions}
\textbf{Magdalayna Curry}: Writing -- original draft, Writing -- review \& editing, Investigation.

\textbf{Minh Duc Chu}: Writing -- review \& editing, Visualization, Software, Methodology, Investigation, Formal analysis, Data Curation, Conceptualization.

\textbf{Changhao Yan}: Writing -- review \& editing, Writing -- original draft.

\textbf{Stuart B. Murray}: Writing -- review \& editing, Supervision, Resources, Investigation, Conceptualization.

\textbf{Kristina Lerman}: Writing -- review \& editing, Writing -- original draft, Supervision, Investigation, Resources, Methodology, Conceptualization.

\textbf{Lindsay E. Young}: Writing -- review \& editing, Writing -- original draft, Supervision, Investigation.


\bibliographystyle{apacite}

\appendix

\section{Supplementary Statistical Tables}
\label{app:stats}

This appendix reports the descriptive and inferential statistics underlying the perceived-harm (Section~\ref{sec:harm}) and engagement (Section~\ref{sec:engagement}) analyses. Perceived harm refers to each video's severity score, computed as the mean of two expert coders' ratings on a 5-point Likert scale (1 = \textit{not harmful} to 5 = \textit{extremely harmful}). Engagement metrics (views, likes, shares, and comments) were recorded at the time of data collection. Because harm ratings are ordinal and engagement counts are heavily right-skewed, group differences were assessed with nonparametric Kruskal--Wallis tests, followed by post-hoc pairwise Dunn's tests (or pairwise Mann--Whitney $U$ tests for the engagement-by-body-type contrasts) with Bonferroni correction for multiple comparisons. $p$-values are reported to three decimal places; values below this precision are reported as $p < .001$.

\begin{table}[ht]
\centering
\caption{Perceived harm to viewers' body image by primary content category. Harm was rated on a 5-point Likert scale (1 = \textit{not harmful}, 5 = \textit{extremely harmful}); each video's score is the mean of two expert coders' ratings. A Kruskal--Wallis test indicated significant differences across categories, $H(4)=40.40$, $p<.001$.}
\label{tab:severity_type_descriptives}
\begin{tabular}{lrrrr}
\toprule
Content category & $n$ & $M$ & $SD$ & $Mdn$ \\
\midrule
Relationship to Body & 438 & 2.44 & 0.87 & 2.50 \\
Relationship to Exercise & 381 & 2.47 & 0.92 & 2.50 \\
Relationship to Food & 303 & 2.70 & 1.06 & 2.50 \\
Relationship to Masculinity & 131 & 2.64 & 1.03 & 2.75 \\
Supplement Abuse & 207 & 2.94 & 1.07 & 3.00 \\
\bottomrule
\end{tabular}
\par\smallskip\raggedright\footnotesize\textit{Note.} $N = 1{,}460$ videos assigned a substantive content category (videos coded as Irrelevant are excluded). $Mdn$ = median.
\end{table}

\begin{table}[ht]
\centering
\caption{Post-hoc pairwise comparisons (Dunn's tests with Bonferroni correction) of perceived harm ratings between primary content categories, following the significant Kruskal--Wallis test reported in Table~\ref{tab:severity_type_descriptives}.}
\label{tab:severity_type_dunn}
\begin{tabular}{lc}
\toprule
Comparison & $p$ \\
\midrule
Body vs. Exercise & 1.000 \\
Body vs. Food & .016\textsuperscript{*} \\
Body vs. Masculinity & .396 \\
Body vs. Supplement Abuse & $<$ .001\textsuperscript{***} \\
Exercise vs. Food & .052 \\
Exercise vs. Masculinity & .685 \\
Exercise vs. Supplement Abuse & $<$ .001\textsuperscript{***} \\
Food vs. Masculinity & 1.000 \\
Food vs. Supplement Abuse & .082 \\
Masculinity vs. Supplement Abuse & .160 \\
\bottomrule
\end{tabular}
\par\smallskip\raggedright\footnotesize\textit{Note.} Category names are abbreviated (e.g., ``Body'' = Relationship to Body). Bonferroni-adjusted $p$-values. \textsuperscript{*}$p<.05$, \textsuperscript{**}$p<.01$, \textsuperscript{***}$p<.001$.
\end{table}

\begin{table}[ht]
\centering
\caption{Post-hoc pairwise comparisons (Dunn's tests with Bonferroni correction) of perceived harm ratings between depicted body types, following a significant Kruskal--Wallis test ($H(3)=84.93$, $p<.001$). Body types were classified with the muscularity figural rating scale (Section~\ref{sec:body_type}).}
\label{tab:severity_bodytype_dunn}
\begin{tabular}{lc}
\toprule
Comparison & $p$ \\
\midrule
Underfat vs. Average & 1.000 \\
Underfat vs. Athletic & .069 \\
Underfat vs. Bodybuilder & $<$ .001\textsuperscript{***} \\
Average vs. Athletic & $<$ .001\textsuperscript{***} \\
Average vs. Bodybuilder & $<$ .001\textsuperscript{***} \\
Athletic vs. Bodybuilder & $<$ .001\textsuperscript{***} \\
\bottomrule
\end{tabular}
\par\smallskip\raggedright\footnotesize\textit{Note.} Harm-relevant videos with a codable body type (videos coded as Irrelevant excluded; $n$: Underfat = 89, Average = 424, Athletic = 613, Bodybuilder = 283). \textsuperscript{*}$p<.05$, \textsuperscript{**}$p<.01$, \textsuperscript{***}$p<.001$.
\end{table}

\begin{table}[ht]
\centering
\caption{Post-hoc pairwise comparisons (Dunn's tests with Bonferroni correction) of engagement metrics between primary content categories. Engagement reflects the cumulative number of likes, shares, comments, and views recorded at the time of data collection. The bottom row reports the global Kruskal--Wallis statistic for each metric.}
\label{tab:engagement_type_dunn}
\begin{tabular}{lcccc}
\toprule
Comparison & Likes & Shares & Comments & Views \\
\midrule
Body vs. Exercise & .059 & .109 & .018\textsuperscript{*} & $<$ .001\textsuperscript{***} \\
Body vs. Food & $<$ .001\textsuperscript{***} & .009\textsuperscript{**} & .214 & .032\textsuperscript{*} \\
Body vs. Masculinity & .005\textsuperscript{**} & .001\textsuperscript{**} & .080 & .069 \\
Body vs. Supplement Abuse & $<$ .001\textsuperscript{***} & .385 & 1.000 & .047\textsuperscript{*} \\
Exercise vs. Food & 1.000 & 1.000 & 1.000 & 1.000 \\
Exercise vs. Masculinity & $<$ .001\textsuperscript{***} & $<$ .001\textsuperscript{***} & $<$ .001\textsuperscript{***} & $<$ .001\textsuperscript{***} \\
Exercise vs. Supplement Abuse & .027\textsuperscript{*} & 1.000 & 1.000 & 1.000 \\
Food vs. Masculinity & $<$ .001\textsuperscript{***} & $<$ .001\textsuperscript{***} & $<$ .001\textsuperscript{***} & $<$ .001\textsuperscript{***} \\
Food vs. Supplement Abuse & 1.000 & 1.000 & 1.000 & 1.000 \\
Masculinity vs. Supplement Abuse & $<$ .001\textsuperscript{***} & $<$ .001\textsuperscript{***} & .013\textsuperscript{*} & $<$ .001\textsuperscript{***} \\
\midrule
Kruskal--Wallis $H(4)$ & 69.78\textsuperscript{***} & 45.12\textsuperscript{***} & 28.20\textsuperscript{***} & 41.54\textsuperscript{***} \\
\bottomrule
\end{tabular}
\par\smallskip\raggedright\footnotesize\textit{Note.} Videos coded as Irrelevant are excluded. Category names are abbreviated (e.g., ``Body'' = Relationship to Body). \textsuperscript{*}$p<.05$, \textsuperscript{**}$p<.01$, \textsuperscript{***}$p<.001$.
\end{table}

\begin{table}[ht]
\centering
\caption{Pairwise Mann--Whitney $U$ tests (Bonferroni-corrected) comparing engagement metrics between depicted body types. Body types were classified with the muscularity figural rating scale (Section~\ref{sec:body_type}); engagement reflects the cumulative number of views, likes, shares, and comments recorded at the time of data collection.}
\label{tab:engagement_bodytype_mwu}
\begin{tabular}{lcccc}
\toprule
Comparison & Views & Likes & Shares & Comments \\
\midrule
Underfat vs. Average & 1.000 & 1.000 & .771 & 1.000 \\
Underfat vs. Athletic & $<$ .001\textsuperscript{***} & $<$ .001\textsuperscript{***} & $<$ .001\textsuperscript{***} & $<$ .001\textsuperscript{***} \\
Underfat vs. Bodybuilder & $<$ .001\textsuperscript{***} & $<$ .001\textsuperscript{***} & $<$ .001\textsuperscript{***} & $<$ .001\textsuperscript{***} \\
Average vs. Athletic & $<$ .001\textsuperscript{***} & $<$ .001\textsuperscript{***} & $<$ .001\textsuperscript{***} & $<$ .001\textsuperscript{***} \\
Average vs. Bodybuilder & $<$ .001\textsuperscript{***} & $<$ .001\textsuperscript{***} & $<$ .001\textsuperscript{***} & $<$ .001\textsuperscript{***} \\
Athletic vs. Bodybuilder & .073 & .064 & .032\textsuperscript{*} & .066 \\
\bottomrule
\end{tabular}
\par\smallskip\raggedright\footnotesize\textit{Note.} Video-level analysis of all videos with a codable body type, including videos coded as Irrelevant ($n$: Underfat = 216, Average = 857, Athletic = 670, Bodybuilder = 297). \textsuperscript{*}$p<.05$, \textsuperscript{**}$p<.01$, \textsuperscript{***}$p<.001$.
\end{table}

\begin{table}[ht]
\centering
\caption{Post-hoc pairwise comparisons (Dunn's tests with Bonferroni correction) of engagement metrics between perceived harm severity categories (Low $\leq 1.5$; Medium $> 1.5$ and $\leq 3.5$; High $> 3.5$). The bottom row reports the global Kruskal--Wallis statistic for each metric.}
\label{tab:engagement_severity_dunn}
\begin{tabular}{lcccc}
\toprule
Comparison & Views & Likes & Shares & Comments \\
\midrule
Low vs. Medium & $<$ .001\textsuperscript{***} & $<$ .001\textsuperscript{***} & $<$ .001\textsuperscript{***} & $<$ .001\textsuperscript{***} \\
Low vs. High & $<$ .001\textsuperscript{***} & $<$ .001\textsuperscript{***} & $<$ .001\textsuperscript{***} & $<$ .001\textsuperscript{***} \\
Medium vs. High & .003\textsuperscript{**} & .036\textsuperscript{*} & .001\textsuperscript{**} & $<$ .001\textsuperscript{***} \\
\midrule
Kruskal--Wallis $H(2)$ & 304.82\textsuperscript{***} & 297.32\textsuperscript{***} & 270.06\textsuperscript{***} & 254.93\textsuperscript{***} \\
\bottomrule
\end{tabular}
\par\smallskip\raggedright\footnotesize\textit{Note.} All videos in the analysis sample, including videos coded as Irrelevant ($n$: Low = 932, Medium = 942, High = 225). \textsuperscript{*}$p<.05$, \textsuperscript{**}$p<.01$, \textsuperscript{***}$p<.001$.
\end{table}

\end{document}